# Hybrid quantum dot-tin disulfide field-effect transistors with improved photocurrent and spectral responsivity

Yuan Huang,[1,a)] Huidong Zang,[1,a)] Jia-Shiang Chen,[1,2] Eli A. Sutter,[3] Peter W. Sutter,[4,b)] Chang-Yong Nam,[1,b)] and Mircea Cotlet[1,2,b)]

[1]*Center for Functional Nanomaterials, Brookhaven National Laboratory, Upton, NY 11973*

[2]*Department of Materials Science and Engineering, Stony Brook University, Stony Brook, NY 11794*

[3]*Department of Mechanical and Materials Engineering, University of Nebraska-Lincoln, Lincoln, NE 68588*

[4]*Department of Electrical and Computer Engineering, University of Nebraska-Lincoln, Lincoln, NE 68588*

We report an improved photosensitivity in few-layer tin disulfide ($SnS_2$) field-effect transistors (FETs) following doping with CdSe/ZnS core/shell quantum dots (QDs). The hybrid QD-$SnS_2$ FET devices achieve more than 500 percent increase in the photocurrent response compared with the starting $SnS_2$-only FET device and a spectral responsivity reaching over 650 A/W at 400 nm wavelength. The negligible electrical conductance in a control QD-only FET device suggests that energy transfer between QDs and $SnS_2$ is the main mechanism responsible for the sensitization effect, which is consistent with the strong spectral overlap between QD photoluminescence and $SnS_2$ optical absorption as well as the large nominal donor-acceptor interspacing between QD core and $SnS_2$. We also find an enhanced charge carrier mobility in hybrid QD-$SnS_2$ FETs which we attribute to a reduced contact Schottky barrier width due to an elevated background charge carrier density.

[a)] these authors contributed equally

[b)] authors to whom correspondence should be addressed. Electronic mail: cotlet@bnl.gov, cynam@bnl.gov, and psutter@unl.edu

Two-dimensional (2D) layered nanomaterials have sparked significant research interest for their potential applications in new types of electronic and optoelectronic devices, including field-effect transistors (FETs), sensors, solar cells, photodetectors, and light emitting diodes.[1-8] Graphene is probably the most popular example of a 2D nanomaterial that has been extensively explored over the years. However, its application in high-performance optoelectronic devices, such as photodetectors[3,9] and photovoltaics[5,10], has been limited due to its zero energy bandgap. Layered metal dichalogenides (LMDs), such as $MoS_2$, $WS_2$, and $WSe_2$, have emerged as alternatives, primarily because of their tunable bandgap energy controlled by thickness (spanning from 1 eV to ~3 eV) and the transition from an indirect bandgap in the bulk to a direct gap in the monolayer,[2,7,11-15] which in turn enhances the 2D material's photoluminescence (PL) quantum yield.[15] Combined with a large surface-to-volume ratio, these material characteristics make LMDs appealing for light harvesting, energy conversion, and chemical sensing device applications. Tin disulfide ($SnS_2$) is a less studied type of LMD, in which the group IV element Sn substitutes the transition metals in other, more familiar LMD compounds such as $MoS_2$ and $WSe_2$. We recently demonstrated FET devices based on mechanically exfoliated $SnS_2$ flakes and achieved high performance on par with those based on $MoS_2$.[14,16] While bulk $SnS_2$ is known to be an n-type semiconductor with an indirect bandgap of 2.2 eV, we found that the bandgap of $SnS_2$ remained indirect even when the layer thickness was decreased down to a single layer, indicating a potentially inferior light absorption cross-section compared with the direct bandgap LMD materials. Recently, hybrid devices combining 2D LMDs with other semiconductors have been explored for the purpose of improving the device performance and functionality.[17-20] Especially for enhancing light absorption and device photosensitivity, the use of colloidal quantum dots (QDs) in a light sensitizing layer in contact with the TMD has been recently proposed.[17,21] QDs have large optical absorption cross-section and wide spectral coverage spanning from ultraviolet to near infrared, depending on their core sizes. Combined with the size-dependent tunable bandgap, such characteristics prove highly effective in the utilization of QDs toward enhancing the optical responsivity of 2D material-based photo-FET devices.[17] We propose that a similar QD-based sensitization scheme should be able to improve the photosensitivity of 2D $SnS_2$ FETs.

Here we report an improved light detection sensitivity in a few-layer $SnS_2$ FET by employing a CdSe/ZnS core-shell QD light sensitization layer. Benefiting from the strong optical absorption of QD as well as the spectral overlap between the PL emission of QDs and $SnS_2$ absorption spectra, which enables an efficient energy transfer from photo-excited QDs to $SnS_2$, QD-$SnS_2$ hybrid FET devices exhibit more than 500 percent increase in the measured photocurrent response with corresponding spectral responsivity greater than 650 A/W at 400 nm wavelength. We find that QD sensitization and light illumination induce an increase in the field-effect electron mobility in the layered $SnS_2$, which we explain based on a reduced contact Schottky barrier width via the increased photo-excited charge carrier density.

Few-layer $SnS_2$ flakes were prepared by mechanical exfoliation from a bulk crystal grown by the vertical Bridgman method and dry-transfer onto a silicon wafer with a 300 nm thick $SiO_2$ dielectric layer, which was also used as a back-gated substrate for FET fabrication and characterization.[14] After locating target $SnS_2$ flakes by bright-field optical microscopy, source-drain contacts (Ti/Au, 10/50 nm) of FETs with 15 μm channel length were fabricated by optical lithography and electron-beam deposition. For the active FET channels, we utilized few-layer $SnS_2$ flakes (~10 monolayers, determined by bright-field optical contrast)[14], instead of monolayer $SnS_2$, because the thicker flakes provide increased light absorption. This enhanced absorption is beneficial not only for the exciton generation within the 2D material but also for the enhanced energy transfer from QDs to $SnS_2$, as observed in our recent single-particle spectroscopy study, where the energy transfer rate was found to increase with increasing number of $SnS_2$ layers.[22]

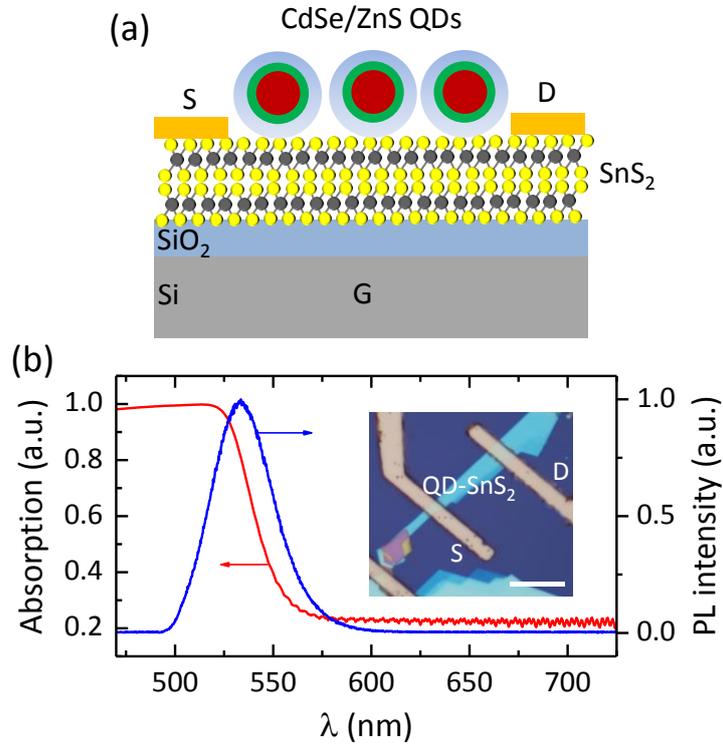

FIG. 1. (a) Schematic depicting a hybrid CdSe/ZnS QD-$SnS_2$ FET device, with S, D, and G denoting source, drain, and gate, respectively. (b) Optical absorption spectrum of a few-layer $SnS_2$ flake (red, left) and PL spectrum of suspended CdS/ZnS QDs in toluene (blue, right). The inset is an optical image of an actual QD-few-layer $SnS_2$ hybrid FET device. Scale bar: 10 μm.

Figure 1(a) shows a schematic of back-gated QD-few-layer $SnS_2$ hybrid FET device with CdSe/ZnS QDs deposited on top of the $SnS_2$ channel via drop-casting from mixed solvents (hexane:octane of 9:1 volume ratio with QD solid concentration of 2.5 mg/l). The $SnS_2$ flake itself exhibits an optical absorption spectrum increasing sharply below ~550 nm (Figure 1(b)), which suggests a bandgap energy of ~2.3 eV. We utilized octadecylamine-coated core/shell CdSe/ZnS QDs with PL spectrum centered at ~535 nm (Ocean NanoTech) to impose a strong overlap with the absorption spectrum of $SnS_2$ (Figure 1(b)). The PL emission peak wavelength of 535 nm is associated with a CdSe core with ~3.4 nm diameter and one monolayer (0.7 nm) thick ZnS shell.[22]

Assuming a thickness of ~2.3 nm for the octadecylamine ligand coating on the QD surface, the interspacing between the CdSe core (edge) and the first layer of the SnS$_2$ flake is thus estimated to be ~3 nm, a distance at which we expect energy transfer to dominate over charge transfer. This large separation between QDs and SnS$_2$, in tandem with the strong spectral overlap between QD PL and SnS$_2$ optical absorption, enables energy transfer from photo-excited QDs to SnS$_2$, as we recently confirmed by time-resolved single-particle PL studies of QD-SnS$_2$ hybrids.[22]

We next measured FET device characteristics in ambient air, under dark and white-light-illuminated (tungsten lamp, ~3.5 mW) conditions. An optical image of the QD-SnS$_2$ hybrid FET device is shown in the inset of Figure 1(b). The SnS$_2$ FET device without QDs and in the dark displays typical n-type drain-source current-voltage ($I_{DS}$–$V_{DS}$) characteristics with increasing $I_{DS}$ toward positive gate voltage, $V_G$. The dark I$_{DS}$ reaches ~8 nA at $V_G$ = 20 V ($V_{DS}$ = 0.5 V), being increased ~10× compared to $I_{DS}$ at $V_G$ = −20 V (Figure 2(a), black curves). Under white light illumination, the photoconductive effect in SnS$_2$ enhances $I_{DS}$ to ~57 nA (at $V_G$ = 20 V, $V_{DS}$ = 0.5 V), which is ~7× larger than the dark $I_{DS}$ obtained under the same conditions (Figure 2(a), red curves). The absolute photocurrent, $\Delta I_{DS,photo} = I_{DS,illuminated} - I_{DS,dark}$, increases from ~10 nA at $V_G$ = −20 V up to ~50 nA at $V_G$ = 20 V, (for $V_{DS}$ = 0.5 V, Figure 2(a) inset). Meanwhile, the FET does not turn off fully under illumination even at $V_G$ = −20 V ($I_{DS}$ = ~14 nA), indicating an elevated background charge carrier density induced by photo-excitation.

For the hybrid FET device with QDs deposited on top of the SnS$_2$ channel, we observed a significant increase in $I_{DS}$ in the dark as well as under white-light illumination. The dark $I_{DS}$ reached 54 nA at $V_{DS}$ = 0.5 V and $V_G$ = 20 V (Figure 2(b), black curves), comparable to the $I_{DS}$ obtained from the bare SnS$_2$ FET (without added QDs) under illumination. At $V_G$ = −20 V, $I_{DS}$ still remains as high as 29 nA, indicating only 1.8× modulation of $I_{DS}$ by a $V_G$ change of −40 V. This suggests an increased background carrier density in QD-SnS$_2$ hybrid FETs even under dark conditions, induced by the application of QDs. We speculate that QDs may introduce a pseudo-gating effect, particularly via their positively charged octadecylamine ligands, which can exert a positive gate electric field onto the QD-SnS$_2$ hybrid FET device. Similar phenomena have been observed in other hybrid systems, such as graphene/polyvinylidene fluoride (PVDF) and 2D MoS$_2$/PbS QDs.[6,17]

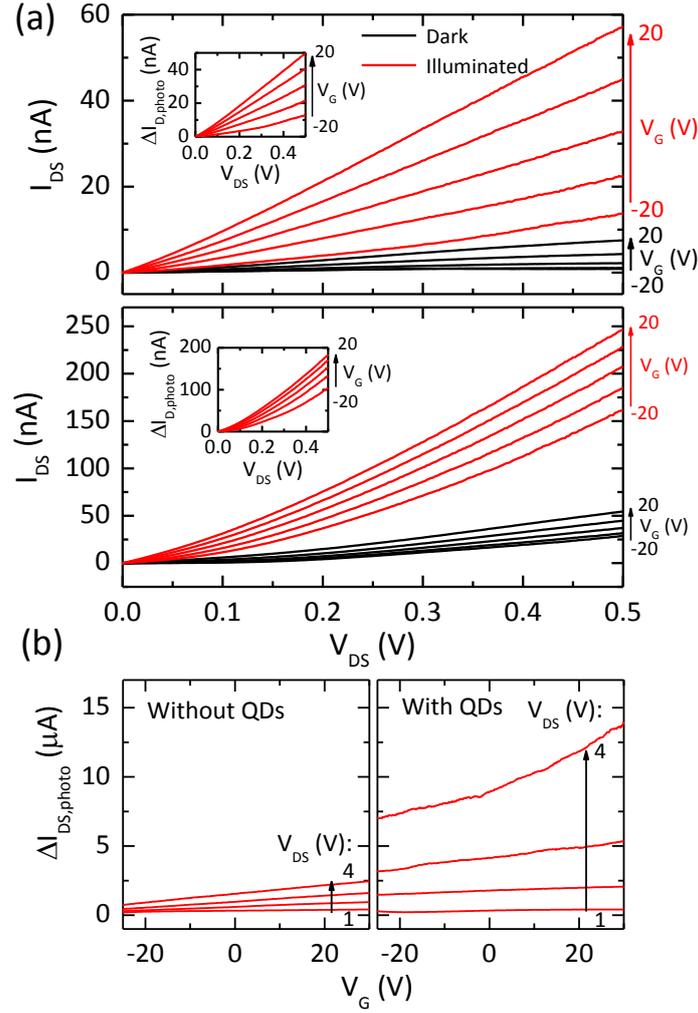

FIG. 2.(a) $I_{DS}$−$V_{DS}$−$V_G$ characteristics of (top) SnS$_2$ FET device and (bottom) QD- SnS$_2$ hybrid FET device in the dark (black curves) and under white light illumination (red curves). $V_G$ is varied from −20 to 20 V, in steps of 10 V. Insets show absolute photocurrent outputs ($\Delta I_{DS,photo}$) with respect to $V_{DS}$ and $V_G$. (b) $\Delta I_{DS,photo}$−$V_G$ characteristics for extended ranges of $V_G$ and $V_{DS}$, for SnS$_2$ FET device (left) and QD- SnS$_2$ hybrid FET device (right). $V_{DS}$ varies from 1 V to 4 V in 1 V increments (indicated by black arrows).

When the hybrid QD-SnS$_2$ FET was exposed to white light, $I_{DS}$ increased to ~250 nA (at $V_{DS}$ =

0.5 V and $V_G$ = 20 V, Figure 2(b), red curves), compared to the dark $I_{DS}$ of 54 nA measured at the same bias condition. Similar to the SnS$_2$-only FET device, the photocurrent output of hybrid QD-SnS$_2$ FET increases for larger $V_{DS}$ and $V_G$ values, with $\Delta I_{DS,photo}$ now reaching up to ~185 nA ($V_{DS}$ = 0.5 V, $V_G$ = 20 V, Figure 2(b) inset), representing a 370% enhancement in the photocurrent output compared to the SnS$_2$-only FET. Even at $V_G$ = −20 V, $\Delta I_{DS,photo}$ for the hybrid FET is as large as ~100 nA, again indicating a significantly elevated background carrier density. Overall, these enhanced photocurrent outputs demonstrate the positive contribution of QDs to the photosensitivity of SnS$_2$ FET devices. A survey of extended ranges of $V_G$ (from −30 V to 30 V) and $V_{DS}$ (from 1 V to 4 V) shows that $\Delta I_{DS,photo}$ of the hybrid QD-SnS$_2$ FET can be further increased to over 10 µA by applying larger $V_G$ and $V_{DS}$. At $V_{DS}$ = 4 V (Figure 2(c)). $\Delta I_{DS,photo}$ of the hybrid QD-SnS$_2$ FET increases from ~7 µA to ~13 µA as $V_G$ changes from −30 V to 30 V, in contrast to a smaller increase of $\Delta I_{DS,photo}$ observed in the SnS$_2$-only FET, from ~0.2 µA to 2.5 µA for the same bias variation. This means that the hybrid QD-SnS$_2$ FET device has >500% enhanced $\Delta I_{DS,photo}$ at $V_G$ = 30V compared with the SnS$_2$-only FET. For the entire range of explored $V_{DS}$ values, $\Delta I_{DS,photo}$ in the hybrid FET is always larger than that in the SnS$_2$-only FET.

The hybrid FET meanwhile cannot be completely turned off even at $V_G$ = −30 V. A similar behavior was reported by Kufer et al. for hybrid PbS QD-MoS$_2$ FETs,[17] where the incorporation of QDs was found to significantly decrease the FET on-off ratio. It is noteworthy that in their demonstration, the authors explained that the enhanced photodetection of MoS$_2$ by the incorporation of PbS QDs was resulting from the *charge transfer* from photo-excited PbS QDs to MoS$_2$. In our study, the insulating shell and ligand coating (giving an estimated spacing between CdSe core and SnS$_2$ of ~3 nm) is believed to prevent such a charge transfer and favor *energy transfer* (via photons, resulting from PL of the QDs and in turn absorbed in the SnS$_2$). The absence of charge transfer can be further confirmed by measuring the electrical conductance of a CdSe/ZnS QD-only FET device (i.e., QDs drop-cast directly onto $SiO_2$, with contact geometry similar to the device shown in Figure 1(b), but without a $SnS_2$ flake). In such a device, we observed a negligible current under both dark and illuminated conditions,[23] proving a negligible charge transfer due to the insulting ZnS shell and ligand coating, and thus confirming that the main interaction mechanism between CdSe/ZnS QDs and $SnS_2$ is energy transfer.

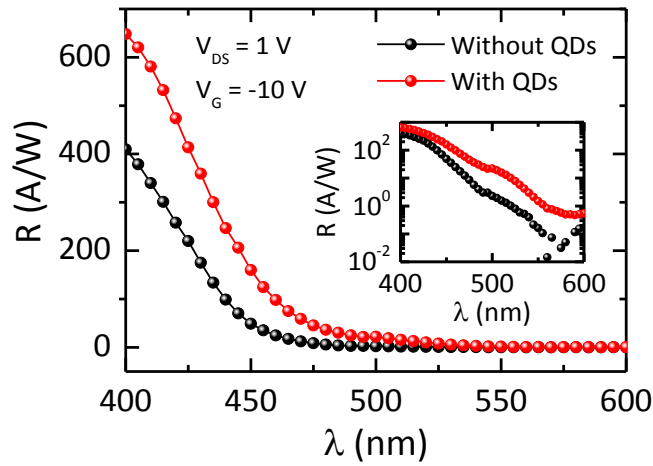

FIG. 3. Spectral responsivity $R$ of a few-layer $SnS_2$ FET device at $V_G = -10$ V and $V_{DS} = 1$ V, before (black symbols) and after (red symbols) the application of the CdSe/ZnS QD sensitization. Inset shows a semi-logarithmic plot of the same data.

We also measured the spectral responsivity, $R = \Delta I_{DS,photo}/P_{light}$ with $P_{light}$ being the incident light power on the device active area, for both the $SnS_2$-only and hybrid QD-$SnS_2$ FET devices in the wavelength ($\lambda$) range from 400 nm to 600 nm, at $V_{DS} = 1$ V and $V_G = -10$ V (Figure 3). Overall, $R$ is enhanced for the hybrid QD-$SnS_2$ FET, reaching ~650 A/W at $\lambda = 400$ nm (c.f., dark $I_{DS}$ ~0.1 nA), compared with 400 A/W for the non-sensitized $SnS_2$-only FET. The onset of $R$ starts at ~550 nm for both FET devices (Figure 3 inset), which is consistent with the absorption band edge of $SnS_2$ (Figure 1(b)). The optical absorption spectrum of CdSe/ZnS QDs in fact also starts at ~550 nm,[24] and thus the measured $R$ after adding QDs mainly features an overall enhanced magnitude with an insignificant change in the spectral shape. It is noted that the onset of $R$ is slow unlike the absorption spectrum (Figure 1(b)), which we suspect to be related to the presence of charge trap states near the band edges of $SnS_2$. Such traps would induce a significant carrier recombination and decreased photocurrent output, consequently lowering the $R$ value near the absorption band edge.

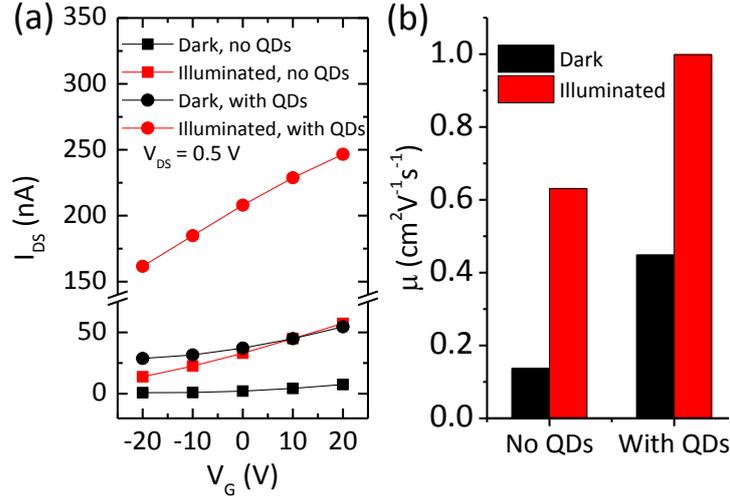

FIG. 4(a) $I_{DS}$−$V_G$ characteristics of few-layer $SnS_2$ FET at $V_{DS}$ = 0.5 V, before (square) and after (circle) CdSe/ZnS QD sensitization, in the dark (black) and under white-light illumination (red). (b) Histogram summarizing the field-effect mobility, $\mu$, calculated from transconductances obtained in (a).

Apart from the optical sensitization, we also found that the hybrid QD-$SnS_2$ FET under illumination exhibits an enhanced field-effect carrier mobility ($\mu$) compared with the few-layer $SnS_2$-only FET. Figure 4(a) shows $I_{DS}$−$V_G$ plot at $V_{DS}$ = 0.5 V, where the FET is not saturated, and thus $\mu$ is proportional to the transconductance ($dI_{DS}/dV_G$) via the following relation:[25] $\mu = (dI_{DS}/dV_G) \times L/(W \cdot C \cdot V_{DS})$ with $L$ and $W$ representing the length (15 μm) and width (~5 μm) of the FET device channel, and $C$ = 11.6 nF/cm$^2$ being the capacitance of the 300 nm thick $SiO_2$ gate dielectric. Figure 4(b) summarizes the obtained $\mu$ values. In the dark, the $SnS_2$-only FET (no QDs) displays $\mu$ = 0.14 cm$^2$V$^{-1}$s$^{-1}$, which is comparable to (but at the low end of) field-effect mobilities reported previously for back-gated few-layer $SnS_2$ FETs at room temperature.[14] Interestingly, $\mu$ increases to over 0.6 cm$^2$V$^{-1}$s$^{-1}$ upon white-light illumination. For the hybrid QD-SnS$_2$ FET, $\mu$ is already higher than 0.4 cm$^2$V$^{-1}$s$^{-1}$ even in the dark, and increases to ~1 cm$^2$V$^{-1}$s$^{-1}$ under white light illumination, which represents a ~7× increase compared with the SnS$_2$-only FET in the dark.

In general, the experimentally measured $\mu$ can be affected by several extrinsic factors, such as contact barrier height, impurity scattering, and temperature. For example, $\mu$ in monolayer MoS$_2$ reported in early studies was lower than 10 cm$^2$V$^{-1}$s$^{-1}$,[26] but it was later shown that the use of graphene contact and h-BN encapsulation could enhance the measured $\mu$ in monolayer MoS$_2$ above 1000 cm$^2$V$^{-1}$s$^{-1}$,[27] and even as high as 34000 cm$^2$V$^{-1}$s$^{-1}$ for six-layer MoS$_2$.[28] This suggests a much higher intrinsic $\mu$ in the 2D material. Given our observed FET device characteristics that show an increased background carrier density during illumination and/or after application of QDs on the SnS$_2$ FET, a potential cause for the observed increase in the measured $\mu$ might be a reduced contact Schottky barrier width $W_D$, a property which is inversely proportional to the carrier density $N$ (i.e., $W_D \propto 1/\sqrt{N}$).[29] Since $N$ affects the FET threshold voltage ($V_T$) according to $V_T \propto \sqrt{N}$,[29] one can expect that $W_D$ decreases with increasing $V_T$ (i.e., $W_D \propto 1/V_T$). From Figure 4(a), we can estimate $V_T$ by extrapolating the linear $I_{DS}-V_G$ plots to $I_{DS} = 0$, and find $V_T$ of ~10 V for the SnS$_2$-only FET in the dark, which increases to ~100 V in the hybrid QD-SnS$_2$ FET under illumination. This 10× increase in $V_T$ implies a tenfold decrease in $W_D$ and thus an easier carrier tunneling through the Schottky barriers at the source and drain electrodes.

In summary, we combined core/shell CdSe/ZnS with few-layer SnS$_2$ to fabricate hybrid QD-SnS$_2$ FETs with improved photo-detection sensitivity via energy transfer from photo-excited QDs to SnS$_2$. Photo-sensitization of SnS$_2$ by the added QDs resulted in an over 5× enhanced photocurrent response in the hybrid FET with the corresponding device spectral responsivity reaching 650 A/W at 400 nm. We also found that the QD sensitization as well as light illumination enhanced the measured carrier mobility in SnS$_2$, which we correlate with an elevated background charge carrier density and consequently decreased contact Schottky barrier width. Our results demonstrate that the energy-transfer-based QD sensitization can be utilized as a new route for enhancing the light harvesting performance of 2D LMD-based opto-electronic devices.

This research was carried out at the Center for Functional Nanomaterials, Brookhaven National Laboratory (BNL), which is supported by the U.S. Department of Energy, Office of Basic Energy Sciences, under Contract No. DE-SC0012704.